\documentstyle[aps,preprint]{revtex}
\tightenlines

\begin{document}
\draft

\title{Nanolithography by non-contact AFM induced local oxidation :\\
Fabrication of tunneling barriers suitable for single electron devices.}

\author{B. Irmer, M. Kehrle, H. Lorenz and J.P. Kotthaus}
\address{Sektion Physik, Ludwig-Maximilians Universit\"at M\"unchen, 80539 
M\"unchen,
 Germany}

\maketitle

\begin{abstract}
We study local oxidation induced by dynamic atomic force microscopy (AFM), 
commonly called TappingMode AFM. This minimizes the field induced forces, 
which cause the tip to blunt, and enables us to use very fine tips. 
We are able to fabricate Ti/TiO$_{x}$ line grids with 18 nm 
period and well defined isolating barriers as small as 15 nm. These junctions
show a non-linear current-voltage characteristic and an exponential dependence 
of the conductance on the oxide width, indicating tunneling as the dominant 
conduction mechanism. From the conductance - barrier width dependence we derive
a barrier height of $\Phi$ = 178 meV.\\
Numerical calculations of the lateral field distribution for 
different tip geometries allow to design the optimum tip for the most localised 
electric field. The electron-beam-deposition (EBD) technique makes it possible 
to actually produce tips of the desired geometry.

\end{abstract}



Proximal probe based lithography has developed over the last years into a 
well established tool for the fabrication of structures and electronic 
devices with nanometric dimensions. Especially the tip-induced oxidation 
- or in more general, tip induced local chemical reactions - have been 
very successful\cite{1}\cite{2}\cite{3}\cite{4}\cite{5} and appear to be 
one of the most promising 
approaches: they preserve the high lateral resolution of the scanned tip
by omitting a subsequent etching step and thus avoid the problem of 
transferring the pattern into an underlying electron system, e.g. metallic 
thin films or heterostructures. Furthermore, it enables one to monitor 
the process in situ by measuring electronic properties, e.g. the 
conductance of a thin channel, defined and constricted by AFM induced 
oxide [6] or the formation of a barrier across a conducting channel \cite{7}.
To optimally use this process in nanofabrication requires (1) the 
understanding of the underlying electrochemical mechanism and the 
parameters that control it, (2) a tip, which is optimised for laterally 
focusing the electric field strength under the experimental conditions 
and (3) a scanning technique which allows one to use these tips and retain 
their properties.\\
Here, we show that operating the AFM in a dynamic, non-contact mode is 
suitable for inducing local oxidation. Hereby the damage to the tip is 
reduced significantly and allows us to address questions involving the 
importance of the tip radius and the overall geometry of the tip.\\ 

We start with thermally oxidised (250 nm SiO$_2$) n-type (10 $\Omega$cm) Si (100) 
material, on top of which 30\AA~to 50\AA~Titanium are thermally evaporated 
at high evaporation rates ($\approx$10~\AA/s) and low background pressure 
(p$\le10^{-8}$mbar).
This metallic film is then patterned  using optical lithography and a HF 
wet etch, finally wire bonded. Local oxidation is performed using a 
commercial AFM (Digital Instruments) and highly doped n$^+$-Si tips 
(NanoSensors), which we additionally sharpen by oxidation. Tip radii 
are well below 100~\AA, typically around 50~\AA. The relative humidity is 
kept constant at 40\% during experiments shown here. The cantilever 
oscillates near its resonance frequency  (approx. 250 kHz) with high 
amplitudes (10-100 nm). The applied tip bias for local oxidation induces 
additional charges on the tip, which bends the cantilever towards the surface. 
This force adds to the normal loading force and can easily damage either 
the tip or the surface. Moreover, in dynamic AFM the force gradient 
$\partial$F/$\partial$z 
due to the electric field changes the force constant $k$ to 
$k^{\star}=k-\frac{\partial F}{\partial z}$, 
shifting the resonance frequency to $\omega_{0}^{\star}=\sqrt{k^{\star}/m}$. The driving bimorph 
oscillates unchanged at the fixed frequency $\omega<\omega_{0}^{\star} < 
\omega_{0}$, and therefore the 
oscillation amplitude is increased. The change in amplitude for a given 
tip bias can easily be measured from amplitude versus distance curves, 
which then can be used to readjust the working setpoint . As the feedback 
is enabled all the time and the damping of the amplitude does not change 
if only the setpoint is readjusted, the overall loading force remains 
unchanged even for applying voltages up to 30~V \cite{7}.
In Fig. 1(a) and 1(b) we show two grids of oxide lines written at a rate 
of 300 nm/sec at a tip bias of -6,5~V. The lines are very regular in width 
(18-20 nm) as well as in height,  even for relatively large scan fields of 
3$\times$3 mm and above. Reducing the period from 120nm to 23nm, parallel conducting 
wires of 6nm linewidth are formed, which are still conductive along the channels 
($\approx$100 k$\Omega$) but isolating in perpendicular direction 
($\gg80$M$\Omega$, at room temperature). 
This demonstrates a complete oxidation process for at least the average of the 
oxide lines. The observed oxide height is 3nm, which agrees very well with what 
is expected from the change in density and molecular weight 
$d_{TiO_2}/d_{Ti}=
\rho_{Ti}/\rho_{TiO_2}\cdot M_{TiO_2}/M_{Ti}\approx 3nm/5nm$. It should be noted that 
at a tapping frequency of $f_{T}=250$ kHz and the observed damping of the amplitude, 
the contact time $t=1/(2f_{T}$) between the tip and the sample surface per cycle is
below $10^{-3}$~ms. As the oscillation amplitude is very large (10-100nm), it is 
unlikely that a stable water meniscus forms between tip and sample. Evidence 
for this is provided by force versus distance curves in contact AFM using the 
same cantilevers (data not shown). If the experiment is to be explained in a 
classical electrochemical set-up, wherein the tip acts as cathode, the water 
film as electrolyte and the sample as anode, the total exposure time is much 
shorter than in contact AFM. However, the total amount of oxidized material is 
very much the same as seen by contact AFM, e.g. by Avouris {\em et al.} \cite{8} for Si, 
or by H. Sugimura {\em et al.} \cite{9} for Ti. We therefore conclude to a corrosion at 
the Ti/TiO$_{x}$ interface, enhanced by the tip-sample electric field in the presence 
of humidity.\\
To define a tunneling barrier we first constrict a predefined 1 mm Ti wire by 
oxidising two large oxide pads, enclosing a 30 nm wide channel (Fig. 2). The 
barrier, perpendicular to the channel, is then oxidized at 2 Hz scanning 
frequency and a tip bias of -4,5 V. In order to avoid the formation of too 
thick barriers with too small tunneling probability by overexposure, we monitor 
the conductance along the channel. As soon as the conductance drops below the 
capacitive signal, oxidation is stopped. The quality of the AFM induced oxide is 
characterized on wide barriers ($\approx$ 100nm). Resistivities of 
$\rho=2\cdot10^{11}~\Omega$cm and max 
field strength $V_{D}=2\cdot10^{6}$ V/cm are measured. These values are similar as for 
macroscopic anodic oxides \cite{10}\cite{11}.\\
At room temperature the devices show an asymmetric, non-linear IV-characteristic. 
This may be understood in the picture of an asymmetrical, shallow barrier, which 
is no longer isotropic for forward and reverse bias. To determine the conduction 
mechanism for these devices, we investigate the dependence of the (tunneling) 
current on the geometrical width of the barrier as obtained from AFM images. 
For four different devices, with barriers varying from 15 nm to 30 nm, the 
current decays exponentially with barrier width, indicating  tunneling as the 
dominant conduction mechanism (Fig.4). Hereof, we obtain a barrier height 
of $\Phi$=178 meV.\\ 
To determine the parameters that affect the lateral resolution of the oxidation 
and therefore to estimate the ultimate limit for this technique, we model the 
lateral field distribution for different tip geometries, namely tip radii and 
cone angles. In a first step we place the tip 10 nm in front of a conducting 
surface. The calculated electric field for a 10 nm sphere (radius in each case), 
10 nm spherical tip with 50$^{\circ}$ pyramidal cone and 5nm spherical tip with 40$^{\circ}$ cone 
are shown in Fig. 5. As expected, the pyramids widen the lateral field compared 
to the free standing sphere, whereas smaller spheres increase the local field 
underneath the  tip. At this stage we do did not consider the growing oxide 
itself as well as the focusing effect of the water layer or meniscus 
because of its large $\epsilon$.
However, for an optimised focusing of the lateral fields, we would like to have a 
needle like tip, which is still sufficient conductive. So called electron beam 
deposited material (EBD) is known to be suitable to define scanning tips with tip 
radii $\le$5nm and very small cone angles \cite{13}. If deposited at high electron energies 
and low beam current densities, they appear to be conductive. Fig. 6 shows an SEM 
image of an EBD tip deposited on top of an NiCr coated Si tip. This tip showes an 
overall resistance R$\le1$M$\Omega$, which is sufficient for applications in electrochemical 
AFM and local oxidation and gives - in contrast to e.g. carbon nanotubes - the unique
possibility to design the tip to exact the requested geometry.\\

In summary, non-contact AFM has been used for locally oxidising Titanium thin films. 
In this mode, the tip-sample forces remain unchanged when applying a tip-sample bias. 
This allows us to use oxide sharpened Si tips, with which we are able to fabricate 
line grids with 6nm structure sizes and 18nm pitch. In situ electrical measurements 
gives fine control over the lithographic process. In this way we fabricated tunneling
barriers as small as 15 nm. The current-voltage caracteristic and the current on 
barrier-width dependence clearly indicates that tunneling is the dominant transport 
mechanism in these devices. Numerical calculations of the lateral distribution of 
the tip to sample electric field indicate an further improvement in the lithographic 
resolution, if only needle like tips with small radii and small cone angles are used. 
We show, that 5nm radius, 5$^{\circ}$ cone angle EBD tips are sufficient conductive to be used 
for local oxidation.\\

The authors would like to acknowledge the contributions of S. Manus and A. Kriele. 
This work was supported financially by the BMBF and by the Volkswagen-Stiftung, which 
we gratefully acknowledge.

\begin{figure}
\caption{Two Ti/TiO$_{x}$ grids written by TappingMode AFM induced local 
oxidation at -6,5V tip bias and a scan speed of 300nm/s. At room 
temperature, the resistance parallel to the lines is 100 k$\Omega$, and 
perpendicular to them $\gg 80 $M$\Omega$.}
\end{figure}

\begin{figure}
\caption{In situ control of the barrier formation. The source-drain conductance 
through the device is monitored while oxidising. The tip is biased at -4V and 
repeatedly scanned at 2Hz across the 30nm wide Ti channel to form a 17nm wide 
barrier.}
\end{figure}

\begin{figure}
\caption{Room temperature current-voltage characteristic of a 20nm wide barrier.} 
\end{figure}

\begin{figure}
\caption{Dependence of the current  on the geometrical barrier width. 4 different 
devices are measured at 100mV bias and T=300 K. The current depends exponentially 
on the barrier width, indicating tunneling as the dominant conduction mechanism. } 
\end{figure}

\begin{figure}
\caption{Calculated electric field distribution for 3 tips with different 
(r,$\Phi$) geometry, 10 nm in front of a conductive plane in vacuum: 10nm sphere, 
pyramidal tip with 10nm radius/50$^{\circ}$ cone angle and 5nm radius/40$^{\circ}$ cone angle, respectively.} 
\end{figure}

\begin{figure}
\caption{SEM image of a EBD tip, deposited onto a commercial Si tip, coated with NiCr. 
Taken from this image, the radius is $\approx$5 nm and the cone angle 5$^{\circ}$. The 
tip shows a resistance of R $\le$1 M$\Omega$, which is sufficient for local oxidation.} 
\end{figure}

\end{document}